\documentclass{article}
\usepackage{spconf,amsmath,amssymb,bm,textcomp,booktabs,graphicx,multirow}

\title{End-to-end DNN based speaker recognition inspired by i-vector and PLDA}                               
\name{Johan Rohdin, Anna Silnova, Mireia Diez, Old\v{r}ich Plchot, Pavel Mat\v{e}jka, Luk\'a\v{s} Burget\thanks{This project has received funding from the European Union’s Horizon 2020 research and innovation programme under the Marie Sklodowska-Curie under grant agreement No. 748097 and the Marie Sklodowska-Curie co-financed by the South Moravian Region under grant agreement No. 665860. The project was also supported by the Google Faculty Research Award program and the Czech Science Foundation under project No. GJ17-23870Y.}}
\address{Brno University of Technology, Brno, Czechia\\ 
\small{\{rohdin, isilnova, mireia, iplchot, matejkap, burget\}@fit.vutbr.cz} }
\begin{document}
\ninept
\maketitle
\begin{abstract}
Recently, several end-to-end speaker verification systems based on deep neural networks (DNNs) have been proposed. These systems have been proven to be competitive for text-dependent tasks as well as for text-independent tasks with short utterances. However, for text-independent tasks with longer utterances, end-to-end systems are still outperformed by standard i-vector + PLDA systems. In this work, we develop an end-to-end speaker verification system that is initialized to mimic an i-vector + PLDA baseline. The system is then further trained in an end-to-end manner but regularized so that it does not deviate too far from the initial system. In this way we mitigate overfitting which normally limits the performance of end-to-end systems. The proposed system outperforms the i-vector + PLDA baseline on both long and short duration utterances. 
\end{abstract}
\section{Introduction}
\label{sec:intro}
In recent years, there have been many attempts to take advantage of neural
networks (NNs) in speaker verification. 
%This is motivated by large performance improvements brought by NNs to many other pattern recognition tasks such as speech recognition \cite{Dahl_taslp_2012} and face recognition \cite{Schroff_2015_CVPR}.
Most of the attempts have replaced or improved one of the components
of an i-vector + PLDA system (feature extraction, calculation of sufficient statistics, i-vector extraction or PLDA) with a neural network. For example by using NN bottleneck features instead of conventional MFCC features \cite{lozano_odyssey_2016}, NN acoustic models instead of Gaussian mixture models for extraction of sufficient statistics \cite{Lei_icassp_2014}, NNs for either complementing PLDA \cite{Novoselov_interspeech_2015,Bhattacharya_SLT16} or replacing it \cite{Ghahabi_icassp_2014}. More ambitiously, NNs that take the frame level features of an utterance as input and directly produce an utterance level representation, usually referred to as an \emph{embedding}, have recently been proposed \cite{Variani_icassp_2014, heighold_icassp_2016, zhang_slt_2016, snyder_slt_2016, Bhattacharaya_interspeech_2017, snyder_interspeech_2017}. The embedding is obtained by means of a \emph{pooling mechanism}, for example taking the mean, over the framewise outputs of one or more layers in the NN \cite{Variani_icassp_2014}, or by the use of a recurrent NN \cite{heighold_icassp_2016}. 
One effective approach is to train the NN for classifying a set of training speakers, i.e., using multiclass training \cite{Variani_icassp_2014, Bhattacharaya_interspeech_2017, snyder_interspeech_2017}. In order to do speaker verification, the embeddings are extracted and used in a standard backend, e.g., PLDA. %Such systems have recently been proven competitive for both short and long utterance durations in text-independent speaker verification \cite{Bhattacharaya_interspeech_2017, snyder_interspeech_2017}. 
Ideally the NNs should however be trained directly for the speaker verification task, i.e., binary classification of two utterances as a \emph{target} or a \emph{non-target} trial \cite{heighold_icassp_2016, zhang_slt_2016, snyder_slt_2016}. Such systems are known as \emph{end-to-end} systems and have been proven competitive for text-dependent tasks \cite{heighold_icassp_2016, zhang_slt_2016} as well as text-independent tasks with short test utterances and an abundance of training data \cite{snyder_slt_2016}. However, on text-independent tasks with longer utterances, end-to-end systems are still being outperformed by standard i-vector + PLDA systems \cite{snyder_slt_2016}. 

%In principle, end-to-end training could be expected to perform better since all parameters of the system are trained jointly for the intended task. 
One reason that end-to-end training has not yet been effective for long utterances in text-independent speaker verification could be overfitting on the training data. 
%Experiments with discriminative training of PLDA suggests that the binary training is prone to overfitting. 
A second reason could be that the previous works have trained the NN on short segments even when long segments are used in testing. This reduces the memory requirements in training and reduces the risk of overfitting but introduces a mismatch between the training and test conditions.

In this work, we develop an end-to-end speaker verification system that is initialized to mimic an i-vector + PLDA baseline.
The system consists of a NN module for extraction of sufficient statistics ({\bf f2s}), an NN module for extraction of i-vectors ({\bf s2i}) and finally, a discriminative PLDA (DPLDA) model \cite{BurgetL_ICASSP:2011,Cumani-pairwise} for producing scores. These three modules are first developed individually so that they mimic the corresponding part of the i-vector + PLDA baseline. After the modules have been trained individually they are combined and the system is further trained in an end-to-end manner on both long and short utterances. During the end-to-end training, we regularize the model parameters towards the initial parameters so that they do not deviate too far from them. In this way the system is prevented from becoming too different from the original i-vector + PLDA baseline which reduces the risk of overfitting. 
Additionally, by first developing the three modules individually, we can more easily find their optimal architectures as well as detect difficulties to be aware of in the end-to-end training.

We evaluate the system on three different data sets that are derived from previous NIST SREs. The three test sets contain speech from various languages and were designed to test the performance both on long (longer than two minutes) and short (shorter than 40s) utterances. The achieved results show that the proposed system outperforms both generatively and discriminatively trained i-vector + PLDA baselines.  

\section{Datasets and Baseline systems}
\label{data}

\subsection{Datasets}

We followed the design of the PRISM\cite{PRISM} dataset in the sense of splitting the data into {\bf training} and test sets. The PRISM set contains data from the following sources: NIST SRE 2004 - 2010 (also known as MIXER collections), Fisher English and Switchboard. During training of the end-to-end system initialization, we used the female portion of the NIST SRE'10 telephone condition (condition 5) to independently tune the performance of the blocks A and B in Figure~\ref{fig:scheme}.

We report results on three different datasets: 
\begin{itemize}
\item The female part of the {\bf PRISM language} condition\footnote{For detailed description, please see section B, paragraph 4 of\cite{PRISM}.} that is based on original (long) telephone recordings from NIST SRE 2005 - 2010. It contains trials from various languages, including cross-language trials. 

\item The {\bf short lang} condition (also containing only female trials) is derived from the PRISM language condition by taking multiple short cuts from each original recording. Durations of the speech in the cuts reflect the evaluation plan for NIST SRE'16 - more precisely we based our cuts on the actual detected speech in the SRE'16 labeled development data. We chose the cuts to follow the uniform distribution:

\begin{itemize}
\item Enrollment between 25-50 seconds of speech
\item Test between 3-40 seconds of speech
\end{itemize}

We split the resulting set into two equally large disjoint sets where speakers do not overlap. We used one part as our {\bf dev} set for tuning the performance of the DPLDA and the end-to-end system. The other part was used for evaluation only. It should be noted that, for simplicity, we test only on single-enrollment trials unlike in our SRE'16 system description where we include multi-enrollment trials \cite{Interspeech2017:Plchot}.

\item Additionally, we report the results on the single-enrollment trials of the NIST SRE'16 evaluation set (both males and females).
\end{itemize}

\subsection{Generative and Discriminative Baselines}
\label{subsec:baseline}
As features we used 60-dimensional spectral features (20 MFCCs, including $C_0$, augmented with their \(\Delta\: \mathrm{and}\: \Delta\Delta\) features). The features were short-term mean and variance normalized over a 3 second sliding window.

Both PLDA and DPLDA are based on i-vectors~\cite{DehakN_TASLP:2011} extracted by means of UBM with 2048 diagonal covariance components. Both UBM and i-vector extractor with 600 dimensions are trained on the {\bf training} set. For training our generative (PLDA) and discriminative (DPLDA~\cite{BurgetL_ICASSP:2011}) baseline systems,  we used only telephone data from the {\bf training} set and we also included short cuts derived from portion of our training data that comes from non-English or non-native-English speakers. The duration of the speech in cuts follows the uniform distribution between 10-60 seconds. The cuts comprise of 22766 segments out of total  85858. Finally, we augmented the training data with labeled development data from NIST SRE'16.

\vspace{0.5em} \noindent\textbf{\bf PLDA:}
We used the standard PLDA recipe, when i-vectors are mean (mean is calculated using all training data) and length normalized. Then the Linear Discriminant Analysis (LDA) is applied prior PLDA training, decreasing dimensions of i-vectors from 600 to 250. We did not perform any additional domain adaptation or score normalization. We also filtered the training data in such a way that each speaker has at least six utterances which reduces it to the total of 62994 training utterances.

\vspace{0.5em} \noindent\textbf{\bf Discriminative PLDA:}
The DPLDA baseline model was trained on the full batch of i-vectors by means of LBFGS optimizing the binary cross-entropy on the training data. We used the {\bf dev} set to tune a single constant that is used for L2 regularization imposed on all parameters except the constant ($k$ in Eq. \ref{DPLDA_scoring}).

All i-vectors were mean (mean was calculated using all training data available) and length normalized. After the mean normalization, we performed LDA, decreasing the dimensionality  of vectors to 250. As an initialization of DPLDA training, we used a corresponding PLDA model. During the DPLDA training, we set the prior probability of target trials to reflect the SRE'16 evaluation operating point (exactly in the middle between the two operating points of SRE'16 DCF\cite{NIST_SRE_2016}).

\section{Proposed end-to-end DNN architecture}
\begin{figure}[t!]
%trim=left botm right top
\centering\includegraphics[clip, trim=3.6cm 0cm 2.5cm 2.7cm, width=1.\linewidth]{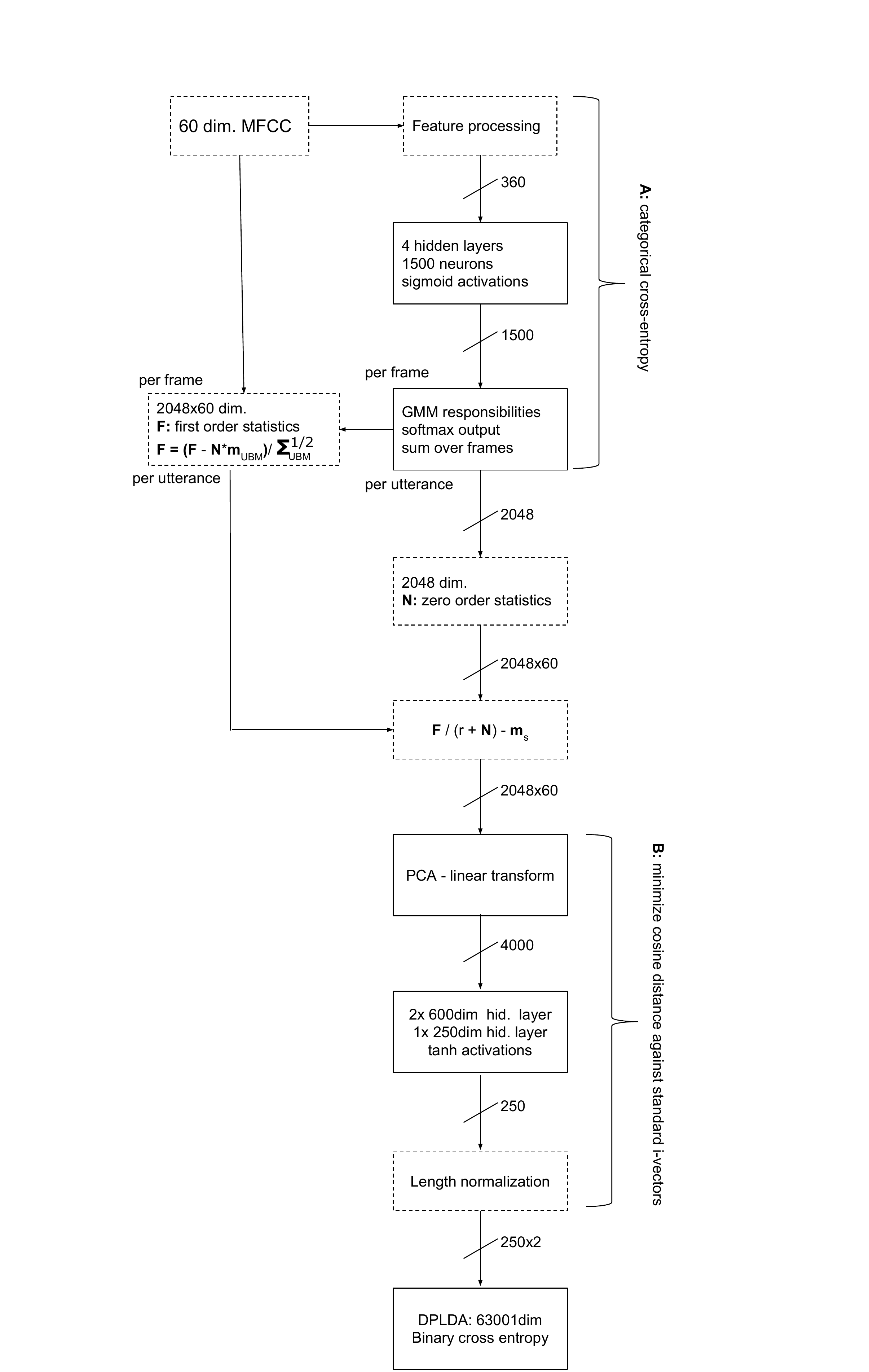}
\caption{Block diagram of the end-to-end system. Part {\bf A} corresponds to the UBM that converts features to GMM responsibilities. By adding the next two blocks we obtain first order statistics ({\bf f2s}). Part {\bf B} ({\bf s2i}) simulates the i-vector extraction followed by LDA and length normalization. Parameters in solid line blocks are meant to be trained, while outputs of the dashed blocks are directly computed.}
\label{fig:scheme}
\end{figure}

In this section, we describe the proposed end-to-end architecture. The system is depicted in Figure  \ref{fig:scheme}.  
We first describe each of the modules \emph{features to statistics}, \emph{statistics to i-vectors} and DPLDA (please see  \cite{arXiv_2017} for details), and then the complete end-to-end system. The system was implemented using the Theano Library \cite{Theano_2016}.
%In Section  \ref{sec:analysis}, we analyse these components more in detail.
\subsection{Features to sufficient statistics}
\label{f2s_sec}
The first module of the end-to-end system converts a sequence of feature vectors into sufficient statistics. We will denote this module as {\bf f2s}. This module consists of a network that predicts a vector of GMM responsibilities (posteriors) for each frame of the input utterance (Block A in Figure \ref{fig:scheme}), followed by a layer for pooling the frames into sufficient statistics. The network that predicts responsibilities consists of four hidden layers with sigmoid activation functions and a softmax output layer. All hidden layers have 1500 neurons while the output layer has 2048 elements which corresponds to the number of components in our baseline GMM-UBM.  We train this network with stochastic gradient descent (SGD) to optimize the categorical cross-entropy with the GMM-UBM posteriors as targets.

As input to the network, the acoustic features described in Section \ref{subsec:baseline} are preprocessed as follows. For each frame, a window of 31 frames around the current frame (i.e. $\pm15\,$ frames) is considered. In this window, the temporal trajectory of each feature coefficient is weighted by a Hamming window and projected into first 6 DCT bases (including $C_0$)~\cite{BUT_babel2014}. This results in a $6\times60=360$-dimensional input to the network for each frame.

Once the network for predicting responsibilities is trained, we add the layer that produces sufficient statistics. 
The input to this layer is a matrix of frame-by-frame responsibilities coming from the previous softmax layer and a matrix of original acoustic features without any preprocessing. This layer is not trained but designed in such a way that it exactly reproduces the standard calculation of sufficient statistics used in i-vector extraction. 
It should be noted that, in principle, expanding the features should not be necessary in order to predict the GMM-UBM posteriors since these are calculated from original features. However, by using the expanded features, we hope that we can gain further improvements in the end-to-end training.

\subsection{Sufficient statistics to i-vectors}
The second module of the end-to-end system is trained to mimic the i-vector extraction from the sufficient statistics (Block B in Figure \ref{fig:scheme}). We will denote this module as {\bf s2i}.
The input sufficient statistics were first converted into MAP adapted supervectors \cite{Reynolds00} (using a relevance factor of $r=16$). To overcome the computational problems that would arise when using the 122880 dimensional supervector as input to the NN, the supervectors were projected by PCA into a 4000 dimensional space. 
The NN consists of two 600 dimensional hidden layers, with hyperbolic tangent (tanh) activation functions. The last layer of the NN is designed to produce length normalized 250 dimensional i-vectors. As training objective, we use the average cosine distance between NN outputs and LDA reduced and length-normalized reference i-vectors. The NN is trained with SGD and L1 regularization.

\subsection{i-vectors to scores (DPLDA)}
\label{subsec:DPLDA}
The final component of the end-to-end system is a DPLDA \cite{BurgetL_ICASSP:2011,Cumani-pairwise} model. The DPLDA model is based on the fact that, given two i-vectors $\bm{\phi}_i$ and $\bm{\phi}_j$, the LLR score for the PLDA model is given by
\begin{eqnarray}
s_{ij}&=&\bm{\phi}_{i}^{T}\bm{\Lambda}\bm{\phi}_{j} +
         \bm{\phi}_{j}^{T}\bm{\Lambda}\bm{\phi}_{i} +
         \bm{\phi}_{i}^{T}\bm{\Gamma}\bm{\phi}_{i} + \bm{\phi}_{j}^{T}\bm{\Gamma}\bm{\phi}_{j} \nonumber \\
&+& (\bm{\phi}_i + \bm{\phi}_j)^T\bm{c} + k,
\label{DPLDA_scoring}
\end{eqnarray}
where the parameters $\bm{\Lambda}$, $\bm{\Gamma}$, $\bm{c}$ and $k$ can be calculated from the parameters of the PLDA model (see \cite{BurgetL_ICASSP:2011} for details). The idea of DPLDA is to train $\bm{\Lambda}$, $\bm{\Gamma}$, $\bm{c}$ and $k$ directly for the speaker verification task, i.e., given two i-vectors, decide whether they are from the same speaker or not. This is achieved by forming trials (usually all possible) from the training data and optimizing, e.g., the binary cross-entropy or the SVM objective. In this work we use the binary cross-entropy objective.

Normally, DPLDA is trained iteratively using full batches, i.e., each update of the model parameters is calculated based on all training data. Whenever the DPLDA model is trained individually in the experiments, we train it in this way. However, for an end-to-end system this would require too much memory and computational time. As is common for neural networks, we therefore calculate each update of the model parameters based on a minibatch, i.e., a randomly selected subset of the training data. 

Due to the fact that the training trials are formed by combining training utterances, it is not obvious how to optimally select the data for minibatches. 
In this paper, we used the following procedure:
\begin{enumerate}
\item 
	\label{enu:mbatch_init}
	For each speaker, randomly group his/her utterances into pairs.\footnote{If a speaker has only one utterance, this utterance will be used as a ''pair''. If a speaker has another uneven number of utterances, one of the pairs will be given three utterances.}
\item
	For each minibatch, randomly select (without replacement) $N$ utterance pairs and use all trials that can be formed from these utterances. If the last pair is selected, repeat Step \ref{enu:mbatch_init}.
\end{enumerate}
This approach limits the number of utterances per speaker in a minibatch. Having more utterances from the same speaker in a batch would give us more target trials but these trials would have been statistically dependent which may affect the training negatively \cite{Rohdin2016}.

\begin{table*}[!th]
\caption{\label{tab:Results}  {Overall results, $C^{\rm Prm}_{\rm min}$ and EER}. Modules marked with a '*' are trained jointly. Other modules are trained sequentially. }
  \centerline{
    %\scalebox{0.9}{
    %\begin{tabular}{l @{} l @{} l @{} l @{} r @{} r @{} r @{} r @{} r @{} r}
    \begin{tabular}{l l  l  l  l  r  r r  r r  r} 
    \toprule
      & \multicolumn{4}{c}{} &   \multicolumn{2}{c}{SRE16} & \multicolumn{2}{c}{short lang} & \multicolumn{2}{c}{PRISM lang}\\
    & System Name & stats & i-vector & PLDA & $C^{\rm Prm}_{\rm min}$  & EER & $C^{\rm Prm}_{\rm min}$  & EER & $C^{\rm Prm}_{\rm min}$  & EER \\
    \midrule \midrule
    1 & Baseline & UBM & i-extractor & Gen.           & 0.988 & 17.645 & 0.699 & 10.303 & 0.411 & 3.902 \\ 
    2 & Baseline DPLDA  & UBM & i-extractor & Discr.  & 0.975 & 16.902 & 0.616 &  9.462 & 0.360 & 3.461 \\ 
    \midrule
    3 & f2s & NN & i-extractor & Gen.                 & 0.980 & 16.809 & 0.687 & 9.866 & 0.394 & 3.713  \\ 
    4 & s2i & UBM & NN & Gen.                         & 0.988 & 16.686 & 0.788 & 11.141 & 0.430 & 4.584\\    
    5 & f2s-s2i & NN & NN & Gen.                       & 0.982 & 16.226 & 0.780 & 11.523 & 0.432 & 4.616\\ 
    6 & f2s-s2i-DPLDA & NN & NN & Discr                & 0.953 & 15.091 & 0.597 & 9.328 & 0.300 & 3.426 \\ 
    \midrule
   7 & s2i-DPLDA\_joint & NN & NN* & Discr.*            & 0.936 & 15.166 & 0.586 & 8.599 & 0.287 & 3.123\\ 
   8 & f2s-s2i-DPLDA\_joint & NN* & NN* & Discr.*       & 0.936 & 15.170 & 0.587 & 8.661 & 0.287 & 3.125\\ 
    \bottomrule  
   \end{tabular}
   %}
  }
\end{table*}

\subsection{End-to-end system}
\label{subsec:e2e}
After the individual components described in the previous subsections have been trained individually, they are combined to an end-to-end system. 
Unfortunately, combining the modules as they are leads to large memory requirements of the end-to-end system. This happens mainly for two reasons. First, contrary to the individual training of the modules, the PCA projection now needs to be part of the network in order for the {\bf f2s} and {\bf s2i} modules to be connected. The PCA matrix with $122880\times4000$ parameters uses approximately 2GB of memory. Second, the {\bf f2s} now needs to process all frames from many different utterances in one batch to obtain the sufficient number of trials for the DPLDA module.
%If the three modules are combined as they are, we can use only approximately 2 utterances per minibatch which is not sufficient for effective training. %(see Section \ref{subsec:mbatch_design} for more discussion about this). 

To mitigate the problem of the large PCA matrix we, before the complete end-to-end training, train only the {\bf s2i} NN and the DPLDA model jointly. As for the individual training of {\bf s2i}, we can use pre-calculated input that includes the PCA projection since this input is fixed as long as {\bf f2s} is not updated. For this training we use minibatches of 5000 pairs ($N$).
To mitigate the large memory requirements of the {\bf f2s} module, we modify the training procedure to keep less intermediate results in memory. Specifically, in usual NN training, the input is first \emph{forward propagated} through the network to get the output of each layer. These outputs are stored in memory and used during \emph{backpropagation} to obtain the derivative of the loss with respect to each model parameter. For the part of {\bf f2s} that calculates responsibilities (Block A in Figure \ref{fig:scheme}), this results in $n_f(1500+1500+1500+1500+2048)$ variables to store in memory, where $n_f$ is the total number of frames. This is much more than in subsequent modules (after pooling the frames into sufficient statistics, $\mathbf{F}$ and $\mathbf  {N}$) where the layer outputs are per utterance. Thus, in order to reduce the memory usage, we calculate the sufficient statistics for one utterance at the time and discard all the layer outputs from Block A once the sufficient statistics for the utterance have been calculated. When the sufficient statistics for all utterances have been obtained, we continue the forward propagation in the normal way, keeping all outputs in memory. During backpropagation, we recalculate the outputs of Block A when needed. This is achieved in a similar way as in \texttt{scan\_checkpoints}\footnote{http://www.deeplearning.net/software/theano/library/scan.html}. This trick allows us to use minibatches of 75 pairs ($N$) instead of approximately 2.  

Unlike the individual training of {\bf f2s} and {\bf s2i}, we use the ADAM optimizer \cite{DBLP:journals/corr/KingmaB14} for training since it may be more robust to different learning rate requirements of the different modules compared to standard SGD. We halved the learning rate whenever we see no improvement in $C^{\rm Prm}_{\rm min}$ on the development set after an epoch (defined to be 250 batches). The training set was the same as for DPLDA.

\section{Results and discussion}
\label{experiments}

We report results in equal error rate (EER) as well as in the average minimum detection cost function for two operating points ($C^{\rm Prm}_{\rm min}$). The two operating points are
the ones of interest in the NIST SRE'16 \cite{NIST_SRE_2016}, namely the probability of target trials being equal to $0.01$ and $0.005$. Table \ref{tab:Results} shows the results for the two baselines, the end-to-end system as well as systems where only some stages of the baseline have been replaced by a NN.
Row 1 and Row 2 show the results for the PLDA and DPLDA baseline, respectively. The DPLDA performs better than generatively trained PLDA on all sets. This is consistent with our previous findings on NIST SRE'16 \cite{Interspeech2017:Plchot}. 

Row 3 shows the results when the UBM is replaced with the {\bf f2s} NN. The i-vector extractor and PLDA model are trained as in the baseline but on the output of the {\bf f2s} NN. It is noticeable that the {\bf f2s} NN performs better than the UBM which it is supposed to mimic. The reason for this seems to be that the {\bf f2s} NN is capable of learning a more robust model that generalizes better to the unseen data than the UBM, mainly because it uses a larger context. Our experiments in the development phase of the {\bf f2s} NN showed that using the 60 dimensional features as input to a 2 layer {\bf f2s} NN gave similar performance as the UBM ($C^{\rm Prm}_{\rm min}$ equal to 0.268 and 0.270 respectively on SRE'10, condition 5) whereas the large context features gave substantial improvement ($C^{\rm Prm}_{\rm min}$ equal to 0.254). 

Row 4 shows the performance when i-vector extractor is replaced by the {\bf s2i} NN. The input is the original statistics from the UBM and a PLDA model is trained on the output. We can see that, except for SRE'16, the performance degrades to some extent compared to the baseline (Row 1).  
%Notice that these results are only for the generatively trained PLDA model. 
Row 5 shows the results when we train a {\bf s2i} module on the output from the {\bf f2s} module instead of the statistics from the UBM. Again, we observe a small degradation compared to using a standard i-vector extractor (Row 3). Interestingly, when we further change from generative trained PLDA to DPLDA, the model performs better than both baselines. This suggests that the output from the {\bf s2i} can well discriminate between speakers but may not well fulfill the PLDA model assumptions so that generative training does not work well. 

After individual training of all blocks, we proceed with joint training of the {\bf f2s} and {\bf s2i} modules, using L2 regularization (tuned on the {\bf dev} set) towards the parameters of the initial models. For this we use a batch size ($N$ in Section \ref{subsec:e2e}) of $5000$ pairs. As can be seen in the Row 7 of Table \ref{tab:Results}, the joint training of the two modules improves the performance on all data sets. Finally, the last row shows the performance when all modules are trained jointly. For this training, we can only use $N=75$ as discussed in Section \ref{subsec:DPLDA}. As can be seen, the performance is almost unchanged from the previous row. There are three possible reasons for this. First, the minibatches might be too small for stable training. Second, with the {\bf f2s} being well initialized and the subsequent modules already being trained to fit its output, the model may be stuck in a local minimum. Third, the {\bf f2s} is in its current design quite constrained. It only estimates the responsibilities but cannot modify the features that are used to calculate the statistics. These issues will be studied in future work. 

In summary, the final system achieved relative improvements with respect to the DPLDA baseline of 3.9\%, 4.7\% and 20.4\% in $C^{\rm Prm}_{\rm min}$ on \emph{SRE16}, \emph{short lang} and \emph{PRISM lang} respectively. In EER, the relative improvements were 10.2\%, 8.5\%, and 9.7\%.

\section{Conclusions and Future Work}
\label{sec:concl}
In this work, we have developed an end-to-end speaker verification system that outperforms an i-vector+PLDA baseline on three different datasets with utterances from many different languages and of both long and short durations. The system was constrained to behave similar to an i-vector + PLDA system. In this way we mitigated overfitting which normally limits the performance of end-to-end systems. This was a conservative approach and future work should explore if less constrained system can perform better, in particular as complement to i-vector+PLDA systems.
We found that joint training of two modules of the three submodules of the system was effective but joint training all three modules was not effective. In future work we therefore want to develop more effective strategies for joint training of all three modules. The proposed system is designed for using single enrollment sessions, and extending it to deal with multiple enrollment sessions is also an important future work.

\bibliographystyle{IEEEbib}
\bibliography{strings,refs}

\end{document}